\documentclass[12pt]{article}

\usepackage{graphicx}

\newcommand{\half}{\textstyle{\frac{1}{2}}}
\newcommand{\dd}{\mathrm{d}}

\begin{document}

\begin{center}
\begin{LARGE}
{\bf The Not-so-simple Pendulum:}

\smallskip
{\bf Balancing a Pencil on its Point}
\end{LARGE}

\smallskip
\begin{small}
{\bf Peter Lynch, UCD, Dublin, May 2014}
\end{small}
\end{center}

\smallskip

{\sc ABSTRACT.}
\begin{small}
Does quantum mechanics matter at everyday scales? We generally assume that its consequences
are confined to microscopic scales. It would be very surprising if quantum effects were
to be manifest in a macroscopic system. This has been claimed for the problem of
balancing a pencil on its tip. The claim has also been disputed. We argue that the
behaviour of a tipping pencil can be explained by the asymptotic properties of the
complete elliptic function, and can be understood in purely classical terms.
\end{small}

\section*{Modelling a balanced pencil}

We have all tried to balance a pencil on its sharp end. Although this is essentially
impossible, you may often notice someone at a tedious committee meeting trying to do it.
We examine the dynamics and see why this unstable equilibrium is unattainable.
We can analyse the pencil using rigid body dynamics, but it is simpler to replace it by
a point mass constrained by a light rod to move in a circle centred at the
point of contact with the underlying surface.

We model the pencil as an inverted simple pendulum with a bob of mass $m$ at one end of a
rigid massless rod of length $\ell$, the other end being fixed at a point. The position of
the bob corresponds to the centre of oscillation of the pencil. We ignore the fine structural
details of the pencil tip, treating it as a point. 

For equilibrium, the bob must be
positioned exactly over the fixed point, and the angular momentum must vanish. In
classical physics, this is theoretically possible. In a quantum system, it is precluded
by the uncertainty principle.

\section*{Dynamics of a simple pendulum}

The story of how Galileo found inspiration in the Cathedral at Pisa is well known.
His mind must have wandered from his prayers as he noticed the regular
oscillations of the chandelier. He concluded, using his pulse to measure the time, that
the period of the back-and-forth swing was constant. Had he been able to measure it more
precisely, he would have realised that the swing-time increases with the amplitude.

Denoting the deflection of the pendulum from the downward vertical by
$\theta$, the dynamical equation is
\[
m\ell\ddot\theta = - mg\sin\theta
\]
where $m$ is the mass of the bob, $\ell$ is the length of the rod and $g$ is the
acceleration due to gravity.  Defining the frequency $\omega = \sqrt{g/\ell}$, this is
\[
\ddot\theta + \omega^2\sin\theta = 0 \,.
\]
For small amplitude motion, we can replace the sine by its argument and the solution is simple
harmonic motion with period $T = 2\pi/\omega$, independent of the amplitude.

For finite --- that is, non-infinitesimal --- motions, the equation is harder to solve but it
is a standard problem in classical dynamics. The solution may be expressed as
\[
\sin\half\theta = \sin\half\theta_0\ \mathrm{sn}[\omega(t-t_0),k]
\]
where $\theta_0$ is the amplitude, $k=\sin\half\theta_0$ and $\mathrm{sn}$ is the Jacobian
elliptic function (see Synge and Griffith, 1959 for full details).  The period is 
\[
T = \frac{4K}{\omega} 
\]
where $K=K(k)$ is the complete elliptic integral of the first kind,
\[
K(k) = \int_0^{\pi/2}\frac{\dd\phi}{\sqrt{1-k^2\sin\phi}} \,.
\]

\begin{figure}[h]
\begin{center}
\includegraphics[scale=0.45]{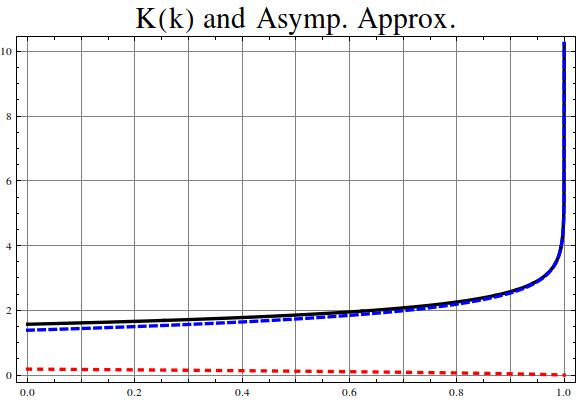}
\caption{Complete elliptic integral of the first kind for $0<k^2<1-10^{-8}$.
Solid: $K(k)$.  Dashed: $\log(4/k^\prime)$. Dotted: Difference, $K(k)-\log(4/k^\prime)$.}
\label{fig:Kfcn}
\end{center}
\end{figure}

For small amplitude --- and therefore small $k$ --- $K$ is approximately $\pi/2$,
so $T \approx {2\pi}/{\omega}$, as in the linear case.
The function $K$ varies very slowly with $k$ so that, for moderate
amplitudes, the period depends only weakly on the value of $\theta_0$. However, $K$ is
unbounded as $\theta_0 \rightarrow \pi$ and the period becomes infinitely long in this limit
(see Fig.~\ref{fig:Kfcn}).

\section*{Asymptotic Estimate of the Period}

There is a solution $\theta\equiv\pi$, independent of time, where the pendulum remains
stationary in an inverted position. This corresponds to a pencil balanced perfectly on
its tip with its centre of mass exactly above the contact point.
However, this solution is unstable and the slightest
disturbance will cause the bob to move away from equilibrium.

The equilibrium implies initial conditions $\theta(0)=\theta_0=\pi$ and
$\dot\theta(0)=\omega_0=0$. 
From the viewpoint of quantum mechanics, we are not free to specify both the
angle $\theta_0$ and angular momentum $p_0=m\ell^2\omega_0$ exactly, 
since they are complementary variables, subject to the Heisenberg uncertainty
principle:
\[
\Delta\theta_0\cdot\Delta p_0 \ge \half\hbar \,.
\]
In the classical case there is no such limitation. However, we will choose initial
conditions such that the deviation of the pendulum from its unstable equilibrium is 
at the quantum scale. 

We set the pendulum parameters to $m=0.1\,$kg and $\ell=0.1\,$m.
Assuming that at $t=0$ the bob is stationary ($p_0=0$) and its displacement from
the vertical is at the atomic scale, we set
\[
\theta_0 = \pi - \delta \qquad\mbox{where}\qquad \delta = 10^{-10}\,\mbox{rad} \,,
\]
and compute the quarter-period using the classical formula $T/4 = K/\omega$.  The
linear displacement of the bob is $\ell\delta = 10^{-11}\,$m, less than an atomic radius.

With $g=10\,$m\,s$^{-2}$ we have $\omega=\sqrt{g/\ell}=10\,$s$^{-1}$ and 
\[
k = \sin\half\theta_0 
  = \sin\left(\displaystyle{\frac{\pi}{2}-\frac{\delta}{2}}\right) 
  = \cos\left(\displaystyle{\frac{\delta}{2}}\right) 
 \approx 1-\displaystyle{\frac{1}{8} \delta^2}
\]
We define $k^\prime$ by $k^2+k^{\prime 2}=1$. 
Then
\[
k^2 \approx 1-\displaystyle{\frac{1}{4} \delta^2}
\qquad\mbox{implies}\qquad
k^{\prime} \approx \displaystyle{\frac{\delta}{2}}
\]

We can use an asympototic expression for the integral $K(k)$.
The \emph{Digital Library of Mathematical Functions} gives the expansion
\[
K(k) = \sum_{m=0}^{\infty} \left[\frac{(\half)_m}{m!}\right]^2 k^{\prime 2m}
\left( \log\left(\frac{1}{k^\prime}\right) + d(m)\right) 
\]
(DLMF, Eqn. 19.12.1. See this reference for full notation).
This expansion is valid for $0<|k|<1$. 
For $k\approx 1$ we can consider only the first term:
\[
K(k) \approx 
\left( \log\left(\frac{1}{k^\prime}\right) + 2\log 2\right)
 = \log\left(\frac{4}{k^\prime}\right) \,.
\]
But since $k^{\prime} \approx \half\delta$, this means
\[
K(k) \approx 
\log\left(\frac{4}{k^\prime}\right) = 
\log\left(\displaystyle{\frac{8}{\delta}}\right) \,.
\]
Using the numerical value $\delta=10^{-10}$, we get $K=25.1$.
Finally, since $\omega=10\,$s$^{-1}$, the quarter-period is
\[
\frac{T}{4} = \frac{K}{\omega} = 2.51\,\mbox{seconds}.
\]

The conclusion is that, even with a tiny deviation of the initial position from
vertical --- less than the width of an atom --- the bob will reach the bottom point
within just a few seconds.

\section*{Solution in Elementary Functions}

We can get an estimate of the time-scale for motion starting near the unstable
equilibrium without using elliptic integrals (Morin, 2008). If $\psi$ is the angle between the
rod and the \emph{upward} vertical, then the motion is governed by
\[
m\ell\ddot\psi = mg\sin\psi \,.
\]
Defining the time scale $\tau = \sqrt{\ell/g}$, the motion near equilibrium, where $\psi$
is small, is described by
\[
\ddot\psi - \left(\frac{1}{\tau^2}\right) \psi = 0
\]
and the solution is
\[
\psi(t) = \half[\psi_0 + \tau v_0] \exp\left({t}/{\tau}\right)
        + \half[\psi_0 - \tau v_0] \exp\left(-{t}/{\tau}\right)
\]
where $\psi_0=\psi(0)$ and $v_0=\dot\psi(0)$. If 
$\psi_0=v_0=0$, we get the (unstable) stationary solution $\psi\equiv 0$.
The negative exponential term is of significance only if the coefficient of the growing
term vanishes. Since we are interested in the growing solution, we drop this second term. 

The initial conditions are constrained by the uncertainty principle
\[
\Delta_x\Delta_p = (\ell\psi_0)\times(m\ell v_0) \ge \half\hbar \,.
\]

We assume that the uncertainties in the dimensionless quantities $\pi-\theta_0$ and
$\omega_0/\omega$ are comparable in magnitude.
This is an arbitrary but reasonable choice, and the results are not sensitively dependent on it.
The coefficient $\half(\psi_0 + \tau v_0)$ is minimum when
the two components are equal. Thus, we set
\[
\psi_0 = \sqrt{\frac{\tau\hbar}{2m\ell^2}}
\quad\qquad\mbox{and}\quad\qquad
v_0 = \sqrt{\frac{\hbar}{2m\ell^2\tau}}
\]
Then the solution becomes
\[
\sqrt{\frac{\tau\hbar}{2m\ell^2}} \exp\left(\frac{t}{\tau}\right)
\]
Since we wish to know how quickly this grows, let us set it equal to unity.
Substituting numerical values $\ell=10^{-1}\,$m and $g=10\,$m\,s$^{-2}$,
we have $\tau = 10^{-1}$s, so that
\[
t = -\frac{\tau}{2}\log\left(\frac{\tau\hbar}{2m\ell^2}\right)
  = -\frac{1}{20}\log\left(\frac{10^{-34}}{2\times10^{-2}}\right) = 3.72\,\mbox{s} \,.
\]
Once again, we find that, even with the smallest disturbances that quantum physics will
allow, the time for the pendulum to swing to the bottom is less than four seconds.

\section*{Discussion}

Are the above results really due to quantum effects? It seems not: in a classical
system, there are no constraints on the initial conditions. There is an equilibrium
solution, corresponding to a perfectly balanced pencil. It is effectively unattainable,
but theoretically possible. In reality, there are always small errors in setting the
initial conditions. What the analysis of the inverted pendulum shows is that, 
however tiny the initial displacement, the pencil will drop within a few seconds.
This is a consequence of the asymptotic properties of the complete elliptic function.
Persistence of balance is impossible in practice.

Morin (2008) interprets the tipping pencil as a macroscopic manifestation of quantum
mechanics. He concludes that `It is remarkable that a quantum effect on a macroscopic
object can produce an everyday value for a time scale.'
However, Easton (2007) concludes that quantum effects are not responsible for
the observed behaviour of pencils, describing this idea as `an urban myth of physics'.
 
\section*{Conclusion}

The above analysis of a balanced pencil can be carried through purely in the context of
classical mechanics, without any reference to the uncertainty principle. In that context,
we are free to choose arbitrary initial conditions. The time to tip is just a few seconds,
even when the deviation of the initial state from the ideal vertical or equilibrium
position is of the order of an atomic radius. This is a consequence of the asymptotic
behaviour of the elliptic integral, which tends logarithmically to infinity as
$k\rightarrow 1$.  The tipping is not a quantum effect, but the result is certainly surprising.


\section*{Sources}
\begin{frenchspacing}

$\bullet$
DLMF: NIST \emph{Digital Library of Mathematical Functions}. Release 1.0.8 of 2014-04-25.
{\tt http://dlmf.nist.gov/19.12.E1}.

$\bullet$
Easton, Don, 2007: The quantum mechanical tipping pencil --- a caution for physics
teachers. \emph{Eur. J. Phys.}, {\bf 28}, 1097--1104.

$\bullet$
Morin, David, 2008:
\emph{Introduction to Classical Mechanics: With Problems and Solutions},
Cambridge University Press. ISBN: 978-0521876223 

$\bullet$
Synge, J. L. and B. A. Griffith, 1959:
\emph{Principles of Mechanics}.
McGraw-Hill, Third Edition, 552pp.

\end{frenchspacing}

\bigskip
\begin{small}
\begin{center}
[Correspondence to {\tt Peter.Lynch :at: ucd.ie}]
\end{center}
\end{small}

\end{document}